# Detection of radiation pressure from solar chameleons

O. K. Baker [1], A. Lindner [2], Y. K. Semertzidis [3], A. Upadhye [4], K. Zioutas [5]

[1] Physics Department, Yale University, PO Box 208120, New Haven, CT 06520, USA;  oliver.baker@yale.edu
[2] DESY, Notkestraße 85, D-22607 Hamburg, Germany;  axel.lindner@desy.de
[3] Brookhaven National Laboratory, NY, PO Box 5000, Upton, NY 11973, USA   yannis@bnl.gov
[4] Argonne National Laboratory, 9700 S. Cass Ave. Lemont, IL 60439, USA;  aupadhye@hep.anl.gov
[5] University of Patras, GR 26504 Patras, Greece;   zioutas@cern.ch

**Abstract:**
*The radiation pressure due to totally reflected (solar) chameleons by thin but dense foils or membranes or mirrors is suggested to be utilized to detect (solar) chameleons and unravel their theoretically motivated behavior, i.e., their coupling to matter ($\beta_m$). The underlying process is distinguished from scattered neutrinos or WIMPs off individual atomic nuclei, which require therefore massive detector targets to compensate for their extremely small interaction cross section with matter. The chameleon reflection is instead a density dependent effect, which distinguishes them theoretically and also experimentally. High sensitive photon radiation pressure or force measuring techniques, which work at the quantum limit, could become the new antennas for chameleons or other particles with similar properties. This may lead to a high detection sensitivity for (solar) chameleons, integrating either over a large focused incident flux, or, measuring eventually chameleons individually. Highly developed opto-mechanical measuring techniques, which already includes gravitational waves antennas, may extend their list of applications to unraveling the nature of dark energy in the Universe, and/or outstreaming of solar chameleons. What looks at first sight as a Gedankenexperiment, might become a real experiment. It is to be remarked that these techniques could be applied to any directed chameleon beam like in a chameleon-through-a-wall experiment for example, in particular when equipped with a Fabry-Pérot resonator.*

## 1. Introduction

Chameleons are particle candidates for the mysterious dark energy (DE) in the Universe [1]. One of their peculiar properties is that their effective rest mass depends on the environment. Thus, when they propagate towards denser matter, which would imply an effective mass becoming larger than their total energy, they would get reflected. The underlying reflection mechanism has been elaborated in [2], providing the theoretical reasoning as well as the conditions under which a reflection in the interface between different dense media should take place.

So far, the detection of chameleons has been based mainly on the inverse Primakoff – effect inside macroscopic magnetic fields, be it in the laboratory be it in space, resembling to a high degree that of axions. In this short note we propose to utilize instead the reflection of chameleons for their detection and identification. We refer throughout this work to chameleons as a generic example, while the same detection principle applies of course to other exotica with similar properties, being motivated theoretically or to be inspired observationally. After all, there is plenty of mysteries in the near or remote outer space, whose origins remain elusive.

## 2. Chameleon phenomenology

As a concrete example of a chameleon theory, we consider a canonically normalized scalar field with the following effective potential $\phi$ in the presence of a nonrelativistic matter density ρ:

$$V_{\text{eff}}(\phi) = M_\Lambda^4 + \frac{M_\Lambda^{n+4}}{\phi^n} + \frac{\beta_m \rho \phi}{M_{\text{Pl}}} \qquad (1)$$

Here, $M_\Lambda = 2.4 \cdot 10^{-3}$ eV is the dark energy scale, and the flat $M_\Lambda^4$ term in $V_{\text{eff}}$ leads to cosmic acceleration without affecting the microphysics. Chameleon theories with such potentials were first proposed and extensively studied in [1, 2]. In a bulk of density ρ, the field rests at the minimum of the effective potential, and the effective mass of the chameleon field is found from the potential at this point:

$$m_{\text{eff}} = V_{\text{eff}}''(\phi)^{1/2} = \sqrt{n(n+1)}\, M_{\text{Pl}} \left(\frac{\beta_m \rho}{n M_{\text{Pl}} M_\Lambda^3}\right)^{\frac{n+2}{2n+2}} \qquad (2)$$

By energy conservation, a chameleon particle with energy E will be unable to enter a region in which $m_{\text{eff}} >$ E. The particle incident upon such a region will reflect [2]. In order to contribute to a radiation pressure, chameleon particles must first enter Earth's atmosphere; thus E > $m_{\text{eff}}(\rho_{atmosphere})$, where $\rho_{atmosphere} = 1.2 \cdot 10^{-3}$g/cm$^3$. Next they must reflect from an object of a typical laboratory density $\rho_{laboratory} \approx$ 10 g/cm$^3$; that is, E < $m_{\text{eff}}(\rho_{laboratory})$. Figure 1 shows that for a chameleon model specified by *n* and $\beta_m$, there is a range of energies between these two masses. For example, if the chameleon field has n = 1 and $\beta_m = 10^6$, then a particle of energy E = 10 eV will cause a radiation pressure. Meanwhile, a particle with E = 0.1 eV will bounce off of the Earth's atmosphere, while a particle with E = 1000 eV will pass right through the laboratory apparatus. Since the mass (2) grows monotonically with ρ for n > 0 as well as for n < −2,

$m_{eff}$ ($\rho_{atmosphere}$) < $m_{eff}$ ($\rho_{laboratory}$) for these potentials, so that a range of energies will always exist in which chameleon particles entering the atmosphere can be reflected by a laboratory apparatus.

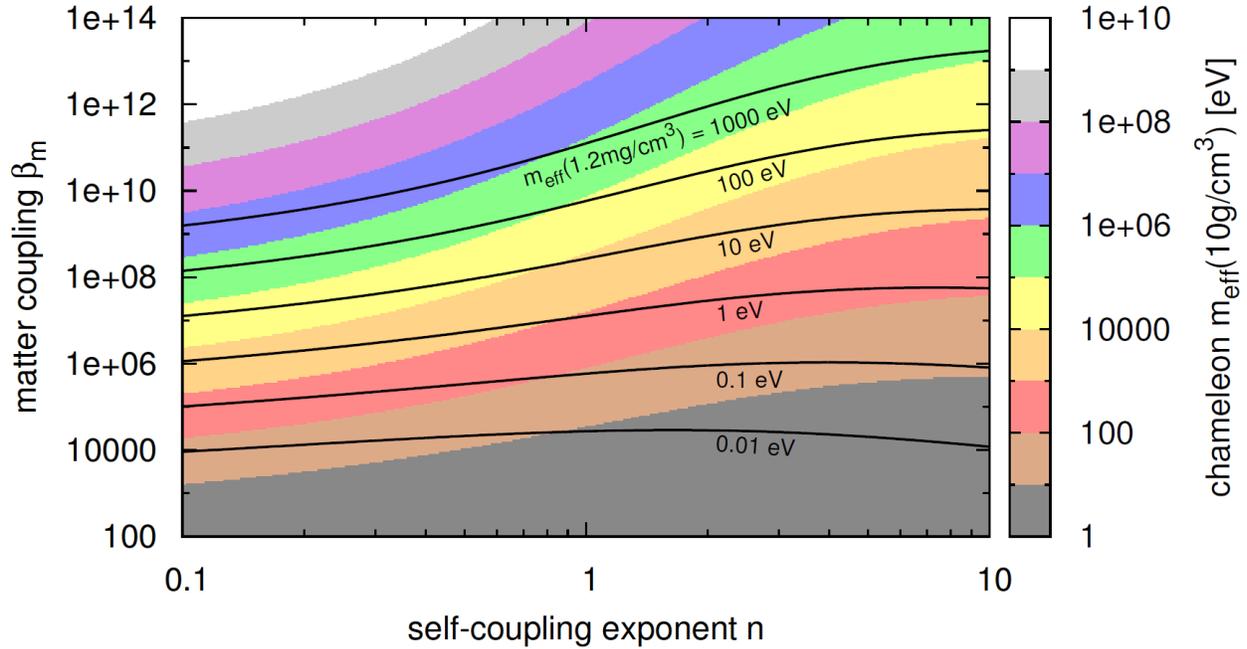

**Figure 1** Chameleon mass as a function of n and $\beta_m$ at densities $\rho_{atmosphere}$ =1.2 mg/cm$^3$ (solid black contours) and $\rho_{laboratory}$ = 10 g/cm$^3$ (shaded regions).

### 3. The detection concept

Reflection of particles like chameleons incident into a denser slab of material results in a momentum transfer, which is similar (in the end effect) to the well known radiation pressure. To be more specific, we assume firstly solar chameleons [3]. So far, it has been estimated to be the only component, which is expected to originate from the tachocline (i.e., at a depth of about 200000 km below the solar surface]. Its spectral shape is a broad distribution peaking at ~600eV [3]. Therefore, it is reasonable to expect that already this component has a low energy tail down to ~10 eV. Following the reasoning of [2], part of this low energy should be reflected by dense matter even at normal incidence, while at smaller grazing angles the maximum energy, which can be reflected, increases. This focusing makes the potential radiation pressure scheme even stronger.

In addition, it is reasonable to assume that lower energy chameleons should be produced similarly (i.e., by the Primakoff – effect) in the magnetic layers above the tachocline. We recall that the ubiquitous solar surface magnetism, and in particular that measured in active regions, i.e., sunspots, reaches 0.1 to 0.5 Tesla. Their origin is widely believed to be deep inside the Sun, i.e., near the tachocline, where magnetic field strengths might be up to 30-100 Tesla [3]. Moreover, following Figure 2 in [4], at a depth below the photosphere surface of only ~5600 km the magnetic field strength is much stronger (~2.5 Tesla) than that measured on the surface. Similarly, Figure 3 in [5] gives a smoothly increasing field all the way down to the tachocline, reaching ~30 Tesla. Therefore, we expect even enhanced lower energetic chameleons emission from the magnetized layers between the tachocline and the Sun's surface, adding thus to the solar chameleon spectrum far below a few 100 eV. A quantitative estimate of this additional solar chameleon component is not as yet available. Though, for the reasoning of the working principle of this proposal, its exact intensity and spectral shape are not crucial.

In fact, the intensity of emitted solar energy, for example, into particles like chameleons can be as much as ~10% of the total solar luminosity [6], without contradicting stellar evolution arguments over the past ~4.5 Gyears. Such an escaping energy is quite large even for every day's experience, without considering even sophisticated measuring devices (see below). This makes the expected radiation pressure, which such an otherwise hidden solar radiation can exert on a very thin but appropriately dense foil / membrane as a promising detection principle, and this is at the heart of the suggestion of this work. In other words, since the theoretically invented chameleons should behave so much differently than all the other Standard Model particles, it is not surprisingly that their detection allows for an equally different experimental approach than we are familiar in dark matter research.

For pedagogical reasons, and before mentioning below some potential techniques with the highest measuring sensitivity, we recall the widely known "light-mill". This is a special kind of radiometer invented in 1873 by the chemist Sir William Crookes [7]. However, more appropriate seems to be the Nichols radiometer, a sensitive torsion balance (1901), which measures the radiation pressure directly [8]. The highly developed opto-mechanical techniques [9] might be proven to be the sensitive devices to adopt for such investigations. Of course, the visible solar light must be eliminated, before a measured torque of the torsion fibre is a measure of the force acting on the mirrors of the torsion pendulum, and this will be the signal for radiation pressure due to reflected chameleon(-like) particles from the Sun. The degree of alignment with the chameleon source, i.e., the Sun, will define the strength of this otherwise hidden radiation pressure, whose diurnal variation can be used for signal identification purposes. With luck, an experiment of this "simplicity", while being built-upon highly sophisticated equipment, could become useful also for such fundamental physics. The question arises as to whether the aforementioned techniques did experience some kind of basal motion in the darkness, but it was not given the appropriate attention. We therefore suggest, driven by experimental curiosity, to follow this first, before elaborating on the properties of exotica like solar chameleons, the mysterious Sun, etc., in order to estimate and predict more precisely the expected chameleon signature.

***Some numerical values***: in order to show the potential of this detection scheme, we assume, for example, the energy flux to be about 1 mW/cm$^2$ of incident solar chameleons, and which have the potential to be reflected. For comparison, the visible solar irradiance is ~136 mW/cm$^2$, which is enormous following every day's experience (in particular during the summer time in Greece). Note, the maximum allowed flux of solar exotica like chameleons, or other particles with similar properties, is about 15 mW/cm$^2$, which, in case of total reflection is equivalent to a radiation pressure of the order of $10^{-8}$ mbar. Note that already in 1903 light pressures smaller than of $10^{-5}$ N/m$^2$ equivalent to $10^{-7}$ mbar have been measured [10]. In fact, a recent optomechanical torsion experiment [11] has shown a radiation pressure sensitivity of about $10^{-11}$ N/m$^2$, which is equivalent to $10^{-13}$ mbar, or about 0.1 μW/cm$^2$. Note, this is already some 5 orders of magnitude below the maximum solar chameleon flux. Furthermore, it is reasonable to assume that nowdays experimental methods (see for example [9,10]), which work at the quantum limit, should perform even better, in particular once they have implemented the peculiar conditions following the reasoning of this work.

Taking into account that an upgraded axion helioscope like CAST [12] could detect solar chameleon fluxes at the mW/cm$^2$ level, this demonstrates the enormous detection potential of the suggested concept in this work. Moreover, with an appropriate configuration design of large area focusing chameleon telescopes [2], e.g., like XMM/Newton, one may envisage an increase of the incident chameleon flux by a factor of about 1000, with the focal plane being adjusted to the measuring mirror or membrane [9]. This implies an improved (solar) chameleon detection sensitivity accordingly.

***Background***:   For the reasoning of this work we have primarily in mind solar chameleons with an energy below ~100 eV, leaving other potential chameleon sources for future work. The inherent physical background of such an experiment will be the electromagnetic radiation from the Sun. In short, this can be either absorbed and/or separated / suppressed by using appropriate lenses transparent to visible and UV radiation, while absorbing higher energetic photons, but reflecting chameleons. A cascade of 3-5 transparent materials with anti-reflection coatings for light will allow to have a "clean" chameleon flux at the $10^{-10}$ level or below, assuming a reflection of the electromagnetic radiation of ~1% or lower. Other experimental conditions might be proven to be more efficient and easier to implement, but it goes beyond the scope of this work to elaborate the experimental configuration details at this stage. For example, instead of blocking the light component, it could be easier to compare pressure and energy flux measurement for photons, assuming that Chameleons only show up in the pressure data.

## 4. Conclusion

In this work we have pointed out the possibility to utilize the radiation pressure exerted by reflected (solar) chameleons by thin but dense materials. For this purpose, a sensitive radiation pressure measuring device, could substantially improve the "conventional" detection scheme using a magnetic helioscope like CAST. With the highly developed sensitive methods to measure radiation pressure, one might envisage the potential to reach even the quantum limit, i.e., to measure the recoil momentum of individually reflected chameleons, which will allow one to reconstruct not only the flux but also the spectral shape of such exotic particles coming from the Sun or beyond. The same or other potential experiments, conducted in weightless conditions in space, might result to higher sensitivity, since the same force does cause much more "visible" effects.

The considered detection mechanism might have implications in outer space. Thus, the Sun's dynamical behaviour, with numerous unexpectedly observed phenomena, should be considered within the described behaviour of out-streaming solar chameleons while they encounter denser layers / fronts, though at much large scales, e.g., flares / CMEs (Coronal Mass Ejections) / solar wind, etc. Likewise, the tracks of satellites might show signatures of extra pressure by solar chameleons, with the observed flyby anomalies deserving further consideration [13]. It remains to be worked out to which extend the Laser Interferometer Space Antenna (LISA) project could detect solar Chameleons for example.

Finally, optomechanical experiments might be the mostly sensitive devices to react on an additional external radiation pressure component like that from solar chameleons.  For example, the measured frequency drifts or the temporally observed irregularities in Figure 5.9 of [14] are of potential interest to be followed further; in particular, in case the observed behaviour resembles solar diurnal behaviour or other solar activity during the time period this measurement was performed. Therefore, such or other experiments being sensitive to external radiation pressure, when performed over longer time periods, i.e. at least one day, a correlation to solar activity should be searched for. Last but not least, the highest sensitivity of Gravitational Waves antennas might be of potential interest, provided their unprecedented sensitive performance is re-considered following the reasoning of this work about an ubiquitous dark radiation pressure.

## Acknowledgments

We thank Mary Tsagri for pointing to the references given in [9] of this work. K. Z. thanks the Physics Department of Yale University, where this work was initiated, for financial support and hospitality, but also for long and inspiring discussions with O.K.B. We also wish to thank the staff of the central CERN library, and in particular Tullio Basaglia and Anne Gentil-Beccot for the continuous support, fast response and patience.